\documentclass[fleqn,usenatbib]{mnras}

\usepackage{newtxtext,newtxmath}

\usepackage[T1]{fontenc}

\DeclareRobustCommand{\VAN}[3]{#2}
\let\VANthebibliography\thebibliography
\def\thebibliography{\DeclareRobustCommand{\VAN}[3]{##3}\VANthebibliography}

\usepackage{graphicx}	
\usepackage{amsmath}	
\defcitealias{2021MNRAS.502.6170C}{Paper~I}
\usepackage{xcolor}

\title[]{Large-scale structures in the stellar wind of fast-rotating stars spawned by the presence of Earth-like planets. }

\author[A. Canet \& A.I. G\'omez de Castro]{
Ada Canet,$^{1,2,3}$\thanks{E-mail: adacanet@ucm.es}
Ana I. Gómez de Castro$^{1,2}$
\\
$^{1}$Joint Center for Ultraviolet Astronomy (JCUVA), Universidad Complutense de Madrid, Plaza de Ciencias 3, 28040, Madrid, Spain\\
$^{2}$Facultad de Ciencias Matemáticas, S.D. Astronomia y Geodesia, Universidad Complutense de Madrid, Plaza de Ciencias 3, 28040, Madrid, Spainn\\
$^{3}$Facultad de Ciencias Físicas, Departamento de Física de la Tierra y Astrofísica, Universidad Complutense de Madrid, Plaza de Ciencias 3, 28040, Madrid, Spain
}

\date{Accepted XXX. Received YYY; in original form ZZZ}

\pubyear{2023}

\begin{document}
\label{firstpage}
\pagerange{\pageref{firstpage}--\pageref{lastpage}}
\maketitle

\begin{abstract}
Forming planets around young, fast-rotating solar-like stars are exposed to an intense X-ray/extreme ultraviolet radiation field and strongly magnetized stellar winds, as a consequence of the high magnetic activity of these stars. Under these conditions, Earth-like exoplanets may experience a rapid loss of their primordial hydrogen atmospheres, resulting in atmosphere-less rocky obstacles for the stellar winds. The interaction of stellar winds with those planets leads to the formation of potentially observable structures due to the formation of large-scale magnetic field and density disturbances in the vicinity of these planets, such as bow shocks, induced magnetospheres and comet-like tails. In this work, we study the interaction between the stellar winds of active, fast-rotating solar-like stars in the superfast-magnetosonic regime with Earth-like, unmagnetized, tenuous atmosphere, planetary obstacles through numerical 3D simulations using the \textsc{pluto} magnetohydrodynamical code. The properties of AB Doradus, a nearby young star with a small rotation period (0.51 days) and a strong flaring activity, have been used to parameterize this early wind state. Bow shock and induced magnetosphere formation are characterized through the alfv\'enic Mach number $M_A$ of the wind, for different stellar wind configurations. Large bow shocks, up to an extension of $\sim$7.0 planetary radii are found for low-$M_A$ winds. The general increase of density, temperature and magnetic field in these large-scale structures formed around planets may result in potentially detectable spectral signatures.
\end{abstract}

\begin{keywords}
MHD -- exoplanets -- stars:winds -- methods:numerical -- shock waves 
\end{keywords}



\section{Introduction}
Young, solar-like stars exhibit faster rotational periods as well as an enhanced magnetic activity in comparison with their more evolved counterparts. The X-ray/extreme ultraviolet (XUV) radiation of the star, responsible of the heating, expansion, and hydrodynamic losses of primary H-rich atmospheres, could reach extremely high values for fast-rotating, magnetically active stars, resulting in XUV fluxes that can be several orders of magnitude greater than the modern Sun \citep{1997ApJ...483..947G,2003A&A...397..147P,2005ApJ...622..680R}. An Earth-like planet evolving under the influence of this radiation may lose its entire primordial hydrogen atmosphere in very short timescales \citep{2015ApJ...815L..12J,2015A&A...576A..87S}. The strong photoionizing radiation may also rapidly ionize the thin layers of the remaining gas around the planet \citep{1997JGR...102.1641A}, suggesting that rocky, atmosphere-depleted/thin-ionosphere obstacles could be a common scenario  around these young, fast-rotating stars. The loss of the primordial hydrogen atmospheres in Earth-like planets supposes indeed an essential requirement for their possible habitability \citep{2015ApJ...815L..12J,2016MNRAS.459.4088O}. 

Beside the influence of the XUV radiation, coronal stellar winds flowing from fast-rotating, magnetically active stars are expected to be faster, more magnetized, hotter and denser in comparison with the current solar wind \citep{2017A&A...598A..24J}. The magnetohydrodynamical (MHD) interaction between these rocky obstacles and the magnetized stellar wind leads to the formation of structures in the vicinity of the planet, in the form of recognizable magnetic and density disturbances in the stellar wind background, including bow shocks, magnetospheres, Alfvén wings and comet-like wakes. The action of these strong winds may produce more extensive and stronger emission structures around the planetary obstacles, resulting in a clear footprint of their presence around young solar-like stars.

The formation of these structures and their properties are dependent both on the properties of the planet (i.e., the presence of an intrinsic magnetic field or atmosphere, the conductivity of the possible present ionosphere, etc.) and the characteristics of the incident stellar wind. The interaction of strong stellar winds and planets with an atmosphere and/or an intrinsic magnetic field has been widely studied in the exoplanetary context, and the formed plasma structures have been predicted and characterized \citep[e.g.,][]{2014ApJ...790...57C,2015A&A...578A...6M,2017AN....338..881D,2018MNRAS.479.3115V,2021MNRAS.507.3626K,2021ApJ...913..130H,2021MNRAS.502.6170C,2022SpWea..2003164V,2022ApJ...934..189C}. However, the formation of these structures have not been characterized in the case of bare, unmagnetized rocky planets, once the primordial hydrogen atmosphere has been almost completely removed. The objective of this work is to fill this gap.\\
Different wind-planet plasma structures are expected in the case of the interaction of strong stellar winds and planets devoid of an atmosphere and with a weak/absent planetary magnetic field, where the interplanetary magnetic field carried with the stellar wind directly interacts with the planetary ionosphere or surface. The ionosphere of the planet results in a highly conductive medium to the highly ionized stellar wind, producing electric currents in the ionospheric cavity around the planet, and leading to the formation of induced magnetospheres, found around the Solar System, in the case of rocky, unmagnetized planets like Mars and Venus \citep{2011SSRv..162..113B}. These structures have been also predicted for non-magnetized hot Jupiters under the influence of strong stellar winds \citep{2017MNRAS.470.4330E}.

Formation of bow shocks around planetary obstacles is a natural consequence of the interaction of these bodies with superfast-magnetosonic stellar winds, also in the case of non-atmosphere, unmagnetized planets. A bow shock constitutes a traveling compressing wave that naturally changes the state of the medium through which it travels, where the wind slows down from super-alfv\'enic velocities to sub-alfv\'enic velocities through the shock. The presence of a bow shock modifies the local properties of the stellar wind, increasing the local density and temperature in in the upstream part of the shock, close to the planetary surface. The terminal location and properties behind the shock has been widely studied for unmagnetized rocky Solar System planets like Mars and Venus, and also in the exoplanetary context \citep[see][hereafter \citetalias{2021MNRAS.502.6170C}, for an study of the interaction of strong winds and Earth-like planets with extended exospheres]{2021MNRAS.502.6170C}. From statistical fitting to bow shock crossing datasets, the main parameters controlling the size of the bow shock are the ratio between the flow velocity and the Alfv\'en velocity, \citep[i.e. the wind Alfv\'en Mach number $M_A$, ][]{1995JGR...100.7907P,2004SSRv..111..233V}, the wind fast-magnetosonic Mach number $M_{ms}$, defined as the ratio between the flow speed and the fast-mode magnetosonic wave speed \citep{1988JGR....93.5461R,2010JGRA..115.7203E,2017JGRA..122..547H,2022JGRA..12730147G} and the extreme ultraviolet flux  \citep[EUV, ][]{2016JGRA..12111474H,2019JGRA..124.4761H}, finding that higher Mach (alfv\'enic and fast magnetosonic) numbers winds and higher solar EUV fluxes produce more compressing (higher density) shocks at closer distances to the planetary surface, in comparison to low-Mach number solar wind.\\
The solar wind at the orbits of the terrestrial planets is characterized by relatively high Mach numbers \citep[$M_A\sim4.5-5.0$ at the Venusian orbit; $M_{ms}$ could variate between values of 4 and 10 at Mars' orbit, being the most frequent value $M_{ms}\sim$8, ][]{2010JGRA..115.7203E}, whose variations corresponds to changes in solar activity. Due to the high magnetic activity of fast-rotating, young stars, afv\'enic and fast mode magnetosonic Mach numbers are expected range among a wide range of values, and then, different bow shocks extensions are expected to be found in the vicinity of planets orbiting these active stars \citep{2005ApJS..157..396E}. \citet{2015ApJ...806...41C} studied the action of strongly magnetized winds in the sub- and the super-alfvénic regime (a minimum alfv\'enic Mach number of 2.9 is considered in the latter case), interacting with an unmagnetized, Venus-like exoplanet. From their simulations, strong shock formation is detected in front of the planet in the super-alfvénic winds, but their location has not been parameterized. Alfvén wings and extended wake formation are detected in the sub-alfvénic cases, but also in the low-$M_A$ configurations, showing the transition between strongly and moderate magnetized winds.

In this study, we address the interaction between a conductive, Earth-like rocky planet, which atmosphere exerts minimum pressure to the ambient stellar wind,  with no intrinsic magnetic field, and the strong winds of a magnetically active, fast-rotating star, limited to the superfast-magnetosonic regime, from low-magnetized ($M_A$=14.0) up to highly magnetized ($M_A$=1.03) winds. The aim of this work is to characterize the possible structures formed around the planetary obstacle under extreme wind conditions, including bow shocks and induced magnetospheres, as a first step for the evaluation of their detectability around young solar-like stars. As a case of application, we have selected the extremely active star AB Doradus A (hereafter AB Dor), a young solar-like star with a rotation period of 0.51 days, often used as a prototype of magnetically active, fast-rotating star. Numerical 3D, single-fluid, magnetohydrodynamical simulations for the interaction of these winds and the Earth-like planet are carried out using the \textsc{pluto} magnetohydrodynamic (MHD) numerical code.\\

This article is structured as follows: in section \ref{sec:winds} we describe the properties of our test star AB Dor, and compute the parameters of their magneto-centrifugally driven winds, analytically obtained for every orbital distance. In section \ref{sec:numerical_setup}, we describe the numerical setup of the simulations carried out in this study. Finally, we report the results and discussion of the properties for the predicted plasma structures in sections \ref{sec:results} and \ref{sec:conclusions}, respectively.

\section{The Stellar Wind of Ab Doradus A}\label{sec:winds}

AB Dor is a K0V rapidly rotating star with an age of about 40–120 Myr \citep{2005ApJ...628L..69L,2011A&A...533A.106G}, stellar radius $R_{\star}=0.96R_{\odot}$ \citep{2011A&A...533A.106G} and mass $M_{\star}=0.86M_{\odot}$ \citep{2006A&A...446..733G}. 

As a consequence of its short rotational period of 0.51 days, AB Dor displays high surface magnetic fields \citep[up to 500G in most active areas, ][]{2007MNRAS.377.1488H}, and other signs of this enhanced magnetic activity like stellar spots \citep{2020A&A...644A..26I}, and frequent phenomena associated with the release of magnetic energy, such as extremely energetic flares detected in X-ray wavelengths \citep{2016IAUS..320..155L}, optical \citep{2019A&A...628A..79S,2021A&A...652A.135S} and UV wavelengths \citep{2002MNRAS.332..409G}, and presumably coronal mass ejections. 

X-ray and extreme ultraviolet (EUV) observations of the corona of AB Dor show hot plasma with temperatures ranging from $\sim$5 up to 20 MK \citep{2003A&A...408.1087S,2005ApJ...621.1009G}, exhibiting a XUV emission that can be several orders of magnitude higher than the Sun. XUV-driven photoevaporation of the H-rich primitive atmospheres of Earth-like planets orbiting an active star like AB Dor, may trigger the complete loss of these gaseous envelopes in a few Myrs, depending on the initial atmosphere density \citep{2015ApJ...815L..12J}.

Due to the fast rotation and the enhanced magnetic activity, the stellar wind of AB Dor is expected to be mainly accelerated by magneto-centrifugal forces \citep{1976ApJ...210..498B}. This results in an increase of several orders of magnitude in the mass loss rate and momentum of the wind with respect to thermally driven Parker winds \citep{1958ApJ...128..664P}, where the wind is accelerated by the action of the outward directed gradient of the gas pressure, resulting in stronger, faster and more magnetized stellar winds for magnetically active fast rotating stars.

To simulate the interaction between the strong stellar winds of an active star and a rocky obstacle, the wind parameters of AB Dor at different stellar orbits have been calculated analytically following the isothermal, one-dimensional \citet[hereafter WD]{1967ApJ...148..217W} solar wind model, scaled for the case of our proxy star. The WD model explicitly includes magneto-centrifugal forces as the main driver of stellar wind acceleration, while also considering the action of pressure gradients in an isothermal plasma. In this sense, other acceleration mechanisms that have proven to contribute the the wind heating and acceleration in the case of active stars, such as heating due to the propagation of Alfvén waves \citep[e.g.][]{2008ApJ...687.1355E,2017ApJ...843L..33G,2020ApJ...896..123S}, have not been considered in this study, isolating the contribution of magneto-centrifugal forces to the wind acceleration. 
One-dimensional stellar wind models constitute a simplification used to obtain wind parameters at any orbital distance, requiring minimal computational resources. Thus, these simplified models are unable to capture the complex 3D magnetic field topology on the stellar surface, that has been proved to have a great impact on the stellar wind properties \citep{2010ApJ...721...80C,2014MNRAS.438.1162V,2015ApJ...798..116R,2016ApJ...832..145R}, including the presence of open-field regions, transverse cooling and heating effects, among other cross-field effects. A comprehensive analysis of stellar winds ejected from active stars needs the thorough examination of all acceleration mechanisms and a detailed and precise information of the stellar magnetic field topology and variability, which is only available for very few sources, falling beyond the scope of this article. Our objective here is to perform a parametric study of the formation of density and magnetic structures, such as bow shocks and induced magnetospheres, around planets whose atmospheres have been almost completely swept away by the combined influence of stellar radiation and winds, while accounting for the magnetic properties of the wind. Here we show that, as a first approximation, the extension and intensity of these structures can be parameterized through the Alfvén Mach number of the wind. This initial approximation serves as a basis for future investigations and more detailed models.\\
Despite the simplification introduced by using 1D models to obtain characteristic wind parameters, they can still be highly useful for studying the interaction between winds and planets, and are commonly used in MHD stellar wind-planet interactions \citep{2017MNRAS.470.4330E,2020MNRAS.498L..53C,2021MNRAS.508.6001C,2021MNRAS.501.4383V,2022MNRAS.509.5858H,2022MNRAS.510.2111K}, especially when considering a broad range of parameters in the stellar corona, as is the case of this study.

In the WD model, the stellar wind solution is obtained assuming  axial symmetry. The steady state solution of the radial velocity of the wind, $v_r$, passes through three critical points: the slow magnetosonic point, the Alfvén point and the fast magnetosonic point, named after the corresponding MHD wave, from closer to further distances to the star, respectively \citep{lamers_cassinelli_1999}. In this sense, the critical surface delimited by the position of each critical point constitute an idealized circumference in the equatorial plane of the star. 
\begin{table}
\centering
\caption{Stellar mass loss rate (low [L, $1\times10^{-13}M_{\odot}/yr$], medium [M, $1\times10^{-12}M_{\odot}/yr$] and high [H, $1\times10^{-11}M_{\odot}/yr$]), and surface equatorial magnetic field $B_{r,eq}$ (low [l, 2.0 G], medium [m, 10.0 G], high [h, 50.0 G] and very-high [l, 200.0 G]) values considered at the base of the wind for the WD solution of the AB Dor stellar wind.
}
\begin{tabular}{ccc}
\hline
\textbf{SW model} & \textbf{$\dot{M}$ ($M_{\odot}/yr$)} & \textbf{$B_{r,eq}$ (G)} \\ 
\hline
\hline

Ll             & $1\times10^{-13}$                             & 2.0              \\
Lm             & $1\times10^{-13}$                             & 10.0            \\
Lh             & $1\times10^{-13}$                             & 50.0             \\
LH             & $1\times10^{-13}$                             & 200.0             \\
\hline
Ml             & $1\times10^{-12}$                             & 2.0              \\
Mm             & $1\times10^{-12}$                             & 10.0             \\
Mh             & $1\times10^{-12}$                             & 50.0             \\
MH             & $1\times10^{-12}$                             & 200.0             \\
\hline
Hl             & $1\times10^{-11}$                             & 2.0              \\
Hm             & $1\times10^{-11}$                             & 10.0             \\
Hh             & $1\times10^{-11}$                             & 50.0          \\
HH             & $1\times10^{-11}$                             & 200.0          \\
\hline
\end{tabular}
\label{tab:table1}
\end{table}

The MHD equations of the WD model were solved using the method fully described in \citet{2005A&A...434.1191P}, also used in \citet{2017A&A...598A..24J} and in \citetalias{2021MNRAS.502.6170C}, where the three critical points of the model are found using an initial guess for the radial velocity at the Alfv\'en point. In order to scale the solar wind WD solutions to our star, the equations require from six initial stellar outputs: the stellar mass, the stellar base wind surface (that is settled slightly above the stellar surface, $\sim1.1R_{\star}$), the rotation period, the base wind temperature, the mass loss rate and the radial component of the surface magnetic field. While the rotation period and the stellar mass and radius of AB Dor are known stellar parameters, we only have estimations for the mass loss rate, the wind temperature and the surface magnetic field of AB Dor, that variate among a wide range of values.

Following the coronal temperatures inferred from the X-ray and EUV observations of AB Dor  \citep{2003A&A...408.1087S,2005ApJ...621.1009G,2013A&A...559A.119L}, we set an average coronal temperature of 10 MK. This value is also in agreement with the empirical model for the rotation-coronal temperature relation given by \citet{2018MNRAS.476.2465O} for solar-like stars, and also with the empirical coronal temperature-$F_X$ flux relation derived in \citet{2015A&A...578A.129J}. Assuming that the base wind temperature linearly scales with the average coronal temperature, we set a temperature at the base of the wind of 7 MK for all our models, considering a scale down factor of 1.36 \citep{2018MNRAS.476.2465O}. The isothermal assumption considered in the WD model leads to excessive values of the temperature at distances far from the stellar surface, and is usual in stellar wind models to use the polytropic approximation. We have opted to evaluate the wind temperature at every orbital distance using the polytropic approach with index $\gamma=1.1$ \citepalias{2021MNRAS.502.6170C}.

For the mass loss rate, we considered values of $1\times10^{-11}, 1\times10^{-12}, 1\times10^{-13}$ $M_{\odot}/yr$ (i.e., from 10 up to 1000 times the solar mass loss rate), following the results from the numerical 3D MHD simulations of the coronal structure of AB Dor carried out in \citet{2010ApJ...721...80C}, where base wind numerical densities of $2\times10^{8}, 1\times10^{9},1\times10^{10} cm^{-3}$ were assumed, after scaling down the coronal plasma density inferred from X-ray observations of AB Dor in order to simulate the "quiet" conditions for the stellar wind. 

The Zeeman–Doppler Imaging maps of the surface magnetic field of AB Dor \citep{2010ApJ...721...80C,2007MNRAS.377.1488H}, show magnetic fields ranging from a few G up to 500 G in the most active areas. As the WD model reduces the stellar wind solution to the stellar equator, we adopted lower magnetic field values of 2.0G, 10.0G, 50.0G and 200.0G for the radial component at the stellar equator, $B_{r,eq}$, considering different magnetic activity scenarios for AB Dor.

All coronal parameter combinations for the calculation of the solution for the stellar wind of AB Dor are collected in \autoref{tab:table1}.\\

\begin{figure}
	\includegraphics[scale=0.56]{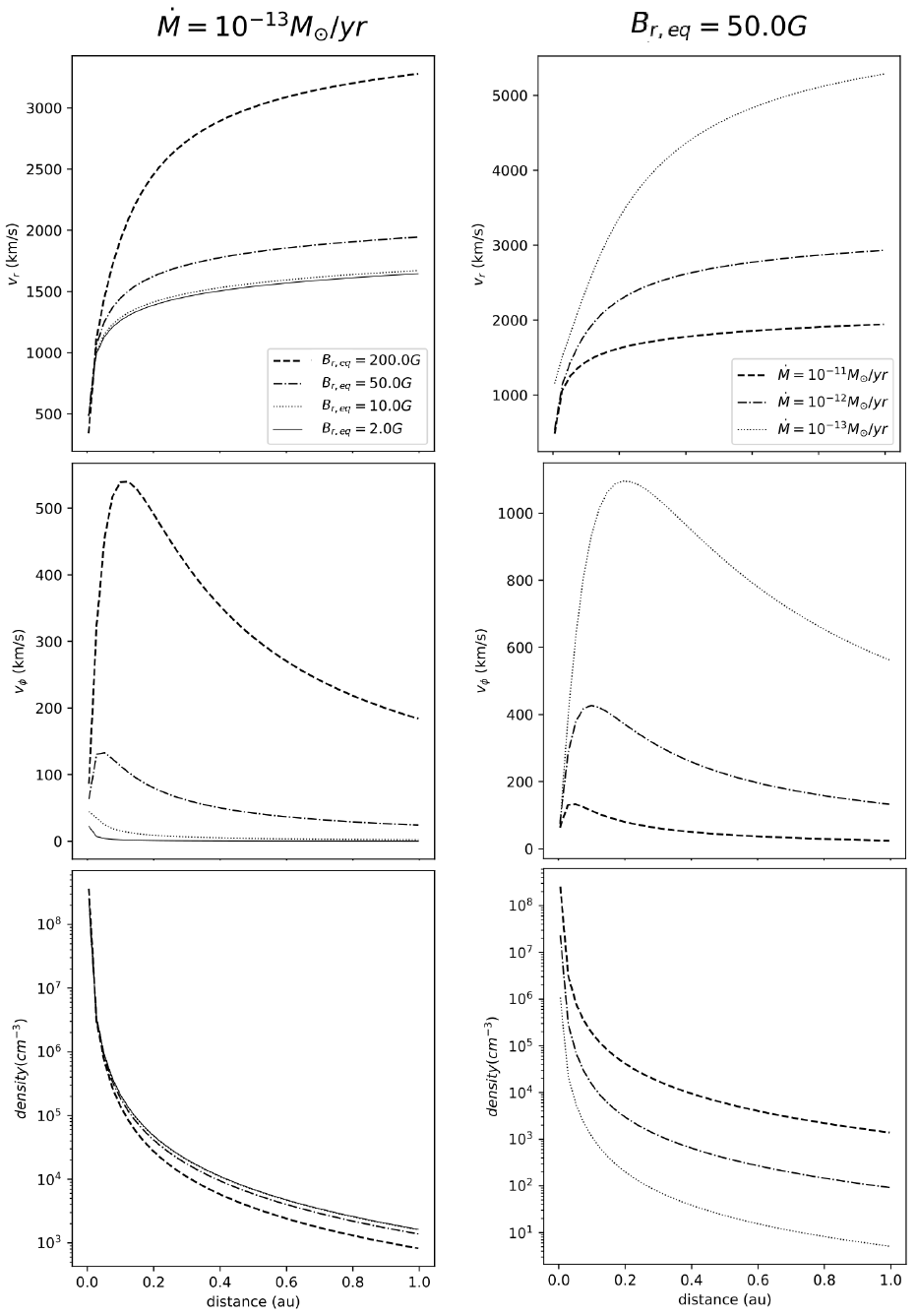}
    \caption{Left column: Radial (top panel), and azimuthal  (middle) velocity components, and density (bottom) profiles at the given distance, solution of the WD stellar wind model for a fixed stellar mass loss rate of $1\times10^{-13} M_{\odot}/yr$, varying the considered equatorial magnetic field $B_{r,eq}$. Right column: Same as left column, for a fixed value of the surface equatorial magnetic field $B_{r,eq}$ of 50.0G, for different values of the mass loss rate.}
    \label{fig:stellar_winds}
\end{figure}

\begin{figure}
	\includegraphics[scale=0.65]{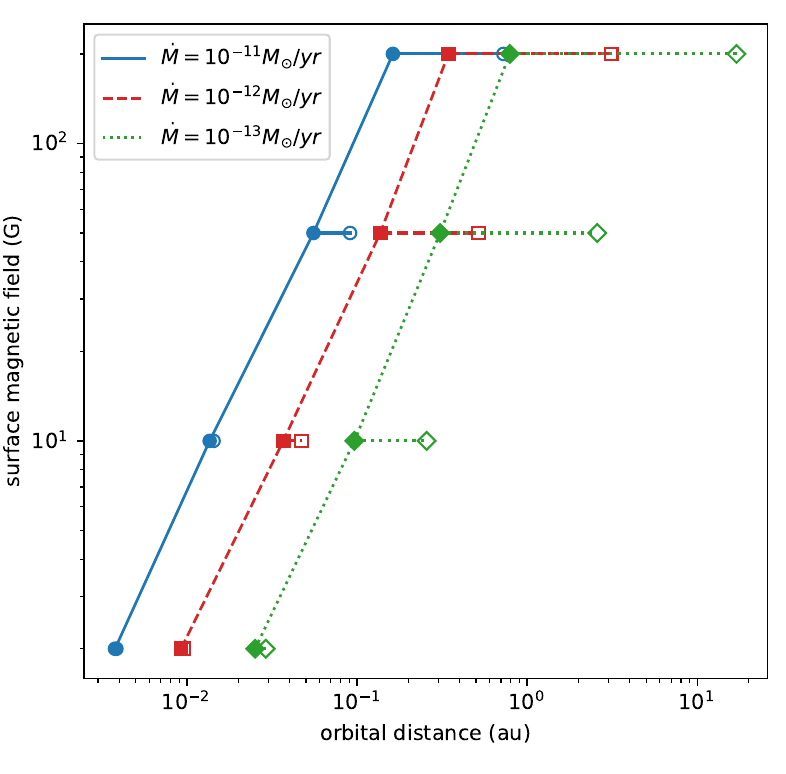}
    \caption{Location of the alfv\'enic (filled markers) and fast magnetosonic (empty markers) critical points, as a function of the surface equatorial magnetic field considered in the model, for each assumed mass loss rate: $1\times10^{-11}$ (solid blue), $1\times10^{-12}$ (dashed red), and  $1\times10^{-13}$ $M_{\odot}/yr$ (dotted green). }
    \label{fig:critical_points}
\end{figure}
The WD analytical solution of the stellar wind of a fast-rotating star like AB Dor shows a clear dependence of the wind parameters at every orbital distance with the surface magnetic field and the mass loss rate of the star. \autoref{fig:stellar_winds} shows how the wind is more accelerated for higher surface magnetic field, while an increase in the considered mass loss rate leads to a decrease of the radial and azimuthal components of the stellar wind speed, as expected from the WD model \citep{1976ApJ...210..498B,1982ApJ...259..180H}. The stellar wind of AB Dor could reach $\sim$10000 km/s at 1 au assuming a magnetic field of 200G and a mass loss rate one order of magnitude higher of that of the current Sun (model LH). The azimuthal component of the wind velocity is significantly lower than the radial component in all cases, being at least one order of magnitude lower in most studied configurations. 

The particle density $n$ at the orbital distance is also affected by the coronal parameters of the star: stellar winds are denser for lower equatorial magnetic fields. Although this result arises as a direct consequence of mass conservation in the WD model, it contradicts solar wind observations, where plasma in regions of closed magnetic field lines (higher magnetic field intensity) is denser compared to the plasma ejected from regions of open magnetic field lines (fast wind), where the plasma is less dense and the magnetic field intensity is lower \citep{1995GeoRL..22.3301P}.  However, this dependence is less critical in comparison to the influence of the mass loss rate (see bottom panels in \autoref{fig:stellar_winds}), where denser stellar winds are found for higher mass loss rates, following again mass conservation. 

The location of the critical points is also sensitive to the mass-loss rate and the surface magnetic field of the star. \autoref{fig:critical_points} shows the Alfv\'en and fast-magnetosonic points location for each mass-loss rate considered in the model, as a function of the radial component of the surface magnetic field $B_{r,eq}$ of the star. Both points are found further away from the stellar surface for higher surface magnetic fields. This trend is also found for lower mass loss rates. The distance between the two points is increased for higher magnetic fields and lower mass loss rates: for low-magnetized stars, the Alfv\'en and the fast-magnetosonic points approximately share the same location, at very close distances to the stellar surface, for each mass-loss rate case. That is, winds are super-alfv\'enic and superfast-magnetosonic from very close distances from the stellar surface in this configuration. On the other extreme scenario, considering the LH model (relatively low mass loss rate and high surface magnetic field), the fast magnetosonic and the Alfv\'en points move forward, at 15 au and 0.7 au, respectively. However, as a result of the use of a simplified 1D stellar wind model, the alfvénic surface appears in the WD model as a perfect circumference around the star, a result that is far from the non-uniform alfvénic surfaces obtained by more sophisticated 3D models accounting for the complex stellar magnetic field topology. Moreover, due to the assumptions regarding the acceleration mechanisms assumed in the WD model, the position of the critical points may differ from those obtained by 3D stellar wind models accounting for other acceleration mechanisms. Nonetheless, the distance at which the critical points are found at the WD model are not relevant in the simulations carried out in the present study, and a wide range of alfvénic Mach numbers have been considered to cover the possible fluctuations of this parameter along the planetary orbit.

These calculations show that the properties of the winds (velocity and density) are strongly dependent on magnetic fields and that a proper study of the wind-planet interaction in very active stars requires using the WD solution instead of the often used Parker thermal wind solution \citep[e.g. ][]{2020MNRAS.498L..53C,2021MNRAS.500.3382C,2022MNRAS.510.3039K}.

The interaction of winds of different nature with an orbiting planet leads to different structure formation around the planetary obstacle. Dayside bow shock formation is expected to form in front of the planet in the case of superfast-magnetosonic winds \citep{2005ApJS..157..396E}, as seen in the case of Solar System planets, where the Alfv\'en point is approximately located at 0.1 au. Wind traveling at sub-alfvénic speeds can magnetically interact with the star, producing angular momentum transfer and modifying the wind topology around the planet \citep{2014ApJ...795...86S,2015ApJ...806...41C}. In this case, the wind is deflected forming the characteristic Alfv\'en wind structure, but no bow shock is formed between the star and the planet.

As we can conclude from the analytical WD solution of the wind of AB Dor, bow shock formation is expected around a planet at almost any orbit for low coronal magnetic fields, while a sub-alfv\'enic interaction is easily produced for more magnetized stars, especially in the case of low mass loss rates.

\section{Numerical Setup}\label{sec:numerical_setup}
\begin{table*}
\caption{Stellar wind parameters for each numerical model of this work. Radial and azimuthal components of the velocity and magnetic field, as well as the characteristic alfv\'enic Mach number $M_A$ are given for a selected orbital distance $d$, from the obtained solutions of the WD analytical model, following the SW model identification given in Table 1.}
\centering
\begin{tabular}{ccccccccc}

\hline
\textbf{model}       & \textbf{SW model}    & \textbf{d (au)}      & \textbf{$v_r$ (km/s)}            & \textbf{$v_{\phi}$ (km/s)} & \textbf{$B_r$ (mG)} & \textbf{|$B_{\phi}|$ (mG)} & \textbf{$M_A$}        & \textbf{$M_{ms}$}\\ \hline

M1                   & Hl                   & 0.05                  & 1105.6  & 4.4                & 20.0                                         & 18.23                                       & 14.0          & 3.6       \\

M2                   & Ll                   & 3.0                  & 1876.6  & 2.2                & 0.005                                        &0.166                                      & 4.6            & 4.0      \\

M3                   & Mm                   & 0.3                  & 1582.5  & 33.2               & 2.78                                         & 10.60                                      & 2.2              & 2.0    \\
M4                   & Hh                   & 0.5                  & 1806.7  & 44.7               & 5.0                                          & 27.9                                       & 1.7           & 1.7       \\
M5                   & Mh                   & 1.2                  & 2972.7  & 117.1               & 0.8                                         & 6.78                                       & 1.2          & 1.1        \\
M6                   & Lm                   & 0.3                  & 2063.0                          & 175.4                                       & 2.78                                         & 7.94                                       & 1.03             & 1.02     \\ \hline

\end{tabular}
\label{tab:table2}
\end{table*}
As we aim to describe the large-scale attributes of the interaction between the magnetized stellar wind plasma and a highly conductive, poor-atmosphere, rocky obstacle, the interaction between the stellar winds of AB Dor and an orbiting planet has been modeled using a single-fluid, ideal MHD numerical code. We performed our simulations in 3D spherical coordinates using the MHD module of the \textsc{pluto} code \citep{2007ApJS..170..228M}. \textsc{pluto} follows a Godunov-type scheme to solve the conservative laws of continuity, momentum, energy, and magnetic induction:
\begin{equation}
    \frac{\partial \rho}{\partial t}+{\nabla} \cdot (\rho \textbf{v})=0 
    \label{masa}
\end{equation}

\begin{equation}
    \frac{\partial(\rho\mathbf{v})}{\partial t} + \nabla\cdot\left[\rho\mathbf{v}\mathbf{v}-\frac{\mathbf{B}\mathbf{B}}{4\pi}+p_T\mathbf{I}\right]^T= \rho(\mathbf{g}+\mathbf{F}_{Cor}+\mathbf{F}_{cen})
\end{equation}

\begin{equation}
    \frac{\partial E_T}{\partial t} + \nabla \cdot \left[(E_T+p_T)\mathbf{v}-\frac{\mathbf{B}}{4\pi}(\mathbf{v}\cdot\mathbf{B})\right] = \rho\left( \mathbf{g}+\mathbf{F}_{cen}\right)\cdot\mathbf{v}
\end{equation}
    
\begin{equation}
    \frac{\partial \mathbf{B}}{\partial t} + \nabla \times \left(\mathbf{B}\times\mathbf{v}\right) = 0
\end{equation}
where $\rho=\mu n m_p$ is the mass density, with $\mu$ the mean molecular weight, $n$ the total number of particles and $m_p$ the proton mass. Here we considered the stellar wind as a fully ionized atomic hydrogenic plasma, so we considered a mean molecular weight $\mu$=0.5. The velocity, magnetic field and gravitational acceleration vectors are denoted as $\mathbf{v}, \mathbf{B}$ and $\mathbf{g}$, respectively. The gravitational acceleration $\mathbf{g}$ includes the acceleration due to the planetary and stellar gravity ($\mathbf{g}$ = $\mathbf{g_{\ast}}+\mathbf{g_{p}}$). The total pressure of the system is $p_T=p+\frac{\mathbf{B}^2}{8\pi}$, including the terms of the thermal ($p$) and magnetic pressures. The total energy of the system, $E_T$, is defined as $E_T=\rho \epsilon + \frac{\rho \mathbf{v}^2}{2}+\frac{\mathbf{B}^2}{8 \pi}$, where $\epsilon$ is the internal energy per mass.\\

An ideal equation of state is solved in conjunction with the ideal MHD equations, using an quasi-isothermal approach with a polytropic index $\gamma$ = 1.1:
\begin{equation}
    \rho\epsilon = \frac{p}{\gamma-1}
\end{equation}

Due to the extreme XUV flux radiation expected in this environment around magnetically active stars, the fully ionization and the quasi-isothermal assumptions considered for the plasma are consistent, as additional adiabatic cooling or heating from ionization are not expected in the distance range considered in the simulations carried out in this work.

To take into account the effects the orbital motion of the planet around the star, we solved the ideal MHD equations in a frame of reference corotating with the planet, i.e., moving with the planet in its Keplerian orbital motion, with a constant angular velocity $\Omega$ equal to the Keplerian orbital velocity of the planet ($\Omega=\sqrt{GM_{\star}/d^3}$, with $d$ the orbital distance from the planet to the center of the star). We modified the right-hand side of the momentum and energy equations, incorporating the corresponding inertial terms of the Coriolis and centrifugal accelerations:
\begin{equation}
    \mathbf{F}_{Cor} = -2(\mathbf{\Omega}\times \mathbf{\vec{v}}) 
\end{equation}
\begin{equation}
    \mathbf{F}_{cen} = -\mathbf{\Omega}\times(\mathbf{\Omega}\times \mathbf{\vec{v}}) 
\end{equation}

Following preliminary simulation tests, the incorporation of centrifugal and Coriolis forces to the MHD equations has no significant effects on the steady state of the simulations in the considered spatial range, due to the strong radial component of the wind velocity in all cases, which dominates the dynamics of the interaction, included those planets orbiting at close orbits to the star.  

The  Harten, Lax, Van Leer (hll) approximate Riemann solver has been selected for numerical flux computation in the interface midpoint. A linear reconstruction associated with a diffusive minmod limiter in all variables is used. A third order Runge-Kutta scheme is implemented in the simulations for time evolution. To ensure the solenoidal condition of the magnetic field, $\nabla \cdot \textbf{B} = 0$, we selected the extended generalized Lagrange Multiplier \citep[GLM, ][]{2010JCoPh.229.5896M,2010JCoPh.229.2117M} formulation, found to be more robust for simulations including low-$\beta$ plasma. \\

The computational grid is made of 384 points in the radial direction $r$, 96 for the polar angle $\theta$, and 192 for the azimuthal angle $\phi$, with $r$ $\in [1, 10R_p]$, $\theta$ $\in[0,\pi]$ and $\phi$ $\in[0,2\pi]$. The corresponding Cartesian coordinate system is made to coincide with the Geocentric Solar Ecliptic System (GSE) of coordinates, i.e., the X-axis direction is pointing from the planet toward the star, the Y-axis is chosen to be in the ecliptic plane and the Z-axis is parallel to the ecliptic pole. The planet is located at the center of the grid, where the maximum resolution of 0.03$R_p$ in the radial, azimuthal and polar coordinates is found. The resolution naturally decreases for increasing radial distance from the center of the grid, with a maximum cell side of of 0.3$R_p$.

Fixed values of density and temperature are set at $r=1.0R_p$ to model the presence of the unmagnetized, tenuous atmosphere planet, of size $1.0R_p = R_{\oplus} = 6.37\times10^{6}m$. We adopted values of $n=500n_{sw}$ for density and a temperature of $T=1\times10^4K$. The calculated thermal pressure corresponding to these values is only $\sim$1.5 times the pressure of the incident stellar wind. In this way, we are able to simulate the interaction of the stellar wind with a planet whose atmosphere exerts a slight pressure on it, ensuring at the same time a constant tenuous wind from the planet's surface. The selected temperature is consistent with the temperature-ionization parameter (defined as $\psi = F_X/(4 \pi n)$) relationship for ionized gases in radiative equilibrium, heated by high-energy photons \citep{2012MNRAS.425.2931O,2016MNRAS.459.4088O}. Considering a X-ray luminosity of $L_X=1\times10^{30}$ erg s$^{-1}$ for AB Dor \citep{2005ApJ...621.1009G}, we calculated for each orbital distance considered in each of the models in this study, assuming a density of $n=500n_{sw}$ in the base of the planetary atmosphere, the ionization parameter is greater than $1\times10^{-4}$ erg cm s$^{-1}$ for each case, approximately corresponding to the saturation temperature of T = $1\times10^4$ K.\\
The planetary surface is modeled as a perfect conductor, i.e., ${B\cdot n = 0}$. Following this approximation, the normal (radial) component of the magnetic field is defined at the ghost cells of the inner boundary with same value but opposite sign as the value of the normal magnetic field in the true cells of the domain (reflective condition). The same condition is defined for the normal component of the velocity, such that the velocity at the planetary surface is $\sim$ 0 km/s. These conditions result in zero magnetic flux across the internal boundary, as
\begin{equation*}
    fx[B_n] = 0.0;
\end{equation*}
\begin{equation*}
    fx[B_t] = V_nB_t - B_nV_t
\end{equation*}
\begin{equation*}
    fx[B_b] = V_nB_b - B_nV_b
\end{equation*}
where the value of normal components is $V_n = B_n = 0$  at the cell interface. Sub-indexes $n$, $t$ and $b$ denote the normal, tangential and bitangential components to the boundary surface. This is consistent with the premise that the remaining atmospheric layers are fully ionized due to the strong photoionizing radiation of AB Dor, not allowing the magnetic field to penetrate inside the planet.

The stellar wind is injected from the outer boundary of
the domain at $r=10R_p$, fixing the stellar wind components of the magnetic
field and velocity vectors, gas pressure and density as constant boundary conditions for $-\pi/2<\phi<\pi/2$, while outflow boundary conditions are defined at $\pi/2<\phi<3\pi/2$. For the considered configurations, the  radial component of the velocity derived from the WD solution is at least one order of magnitude higher than the azimuthal component (see \autoref{tab:table2}). For this reason, in our numerical setup we match the radial components of the velocity and magnetic field with the
X axis of the GSE coordinate systems, while the azimuthal component corresponds to the GSE Y coordinate. In all the models considered, the azimuthal component of the magnetic field exceeds by at least 4 times the radial component, with the exception of model M1. Thus, the magnetic field is approximately perpendicular to the speed of propagation of the wind under these configurations. However, both components of magnetic field are considered in our simulations for studying the possible change in the morphology of the formed MHD structures around the planet.\\ 
The wind thermal pressure $p$ is defined at the
left boundary condition as $p = c_s^2\rho/\gamma$, where $c_s$ is the corresponding
sound speed of the wind.

Here, we performed a grid of six simulation
models, studying the interaction of winds in the superfast-magnetosonic regime, for different Mach ($M_{A}=|\textbf{v}|/|\textbf{A}|$), with $\textbf{A} = \textbf{B}/\sqrt{4\pi\rho}$ the alfv\'enic velocity, and superfast-magnetosonic number ($M_{ms} = |\textbf{v}|/\sqrt{c_s^2+|\textbf{A}|^2}$), ranging from $M_{A} = 14.0$ up to $M_{A} = 1.03$. The parameters for each wind model are obtained from the WD analytical solution for the stellar wind (see second column in \autoref{tab:table2}), calculated for various coronal properties, are obtained at different orbital distances, from 0.05 au to 1 au. This grid of models has been defined to cover a wide range of Alfv\'en Mach numbers since these determine the type of MHD structures formed in the wind (see below).\\

We initialized the simulations defining the stellar wind parameters in the whole domain, except at the internal boundary, where conditions are fixed as described above.

\section{Results}\label{sec:results}
\begin{figure*}

	\includegraphics[scale=0.57]{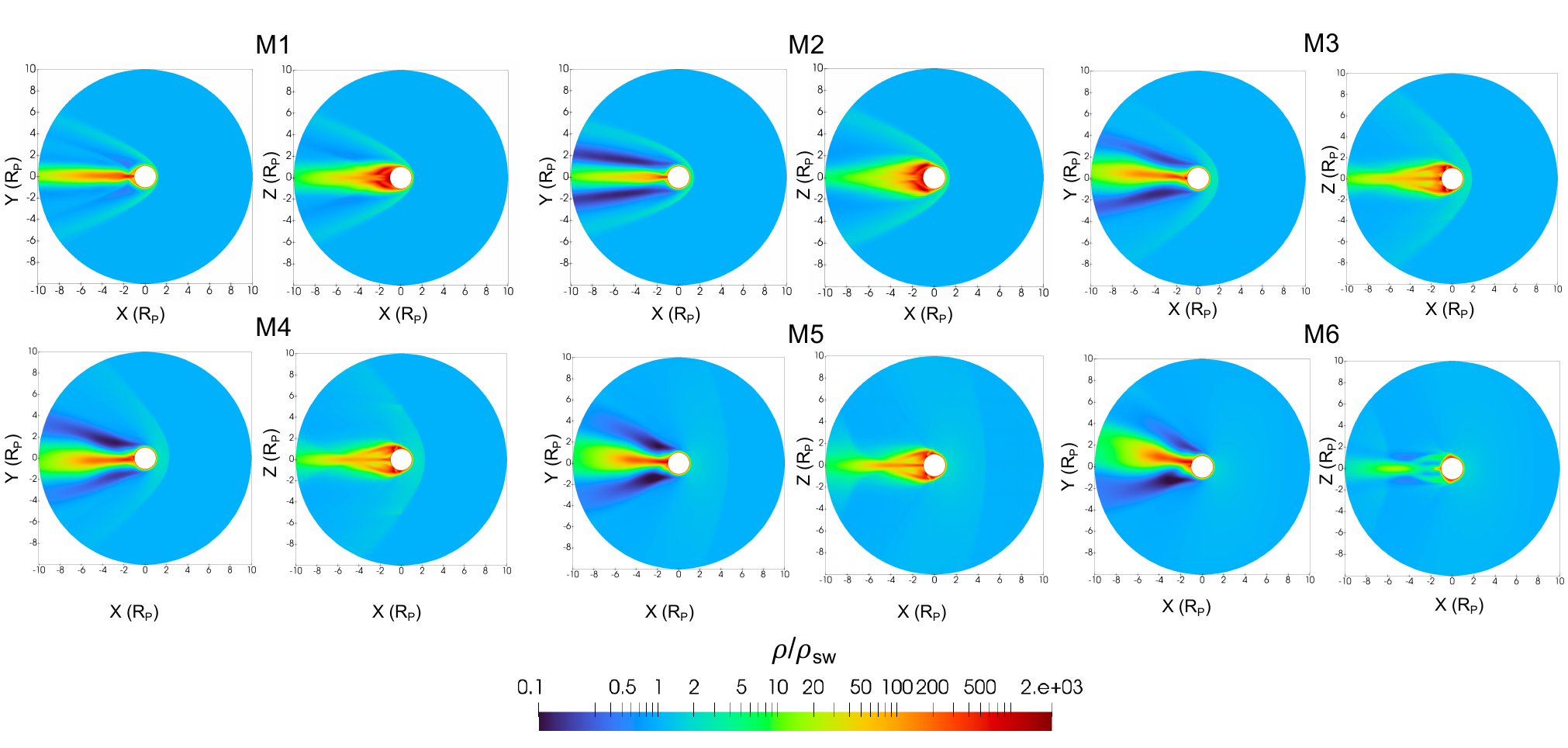}
    \caption{Local distribution of density plasma around the planetary obstacle for each simulated model. The density is normalized by the interplanetary density value, defined as injected in the right boundary side of the domain.}
    \label{fig:mosaic_dens}
\end{figure*}

\begin{figure}

	\includegraphics[scale=0.56]{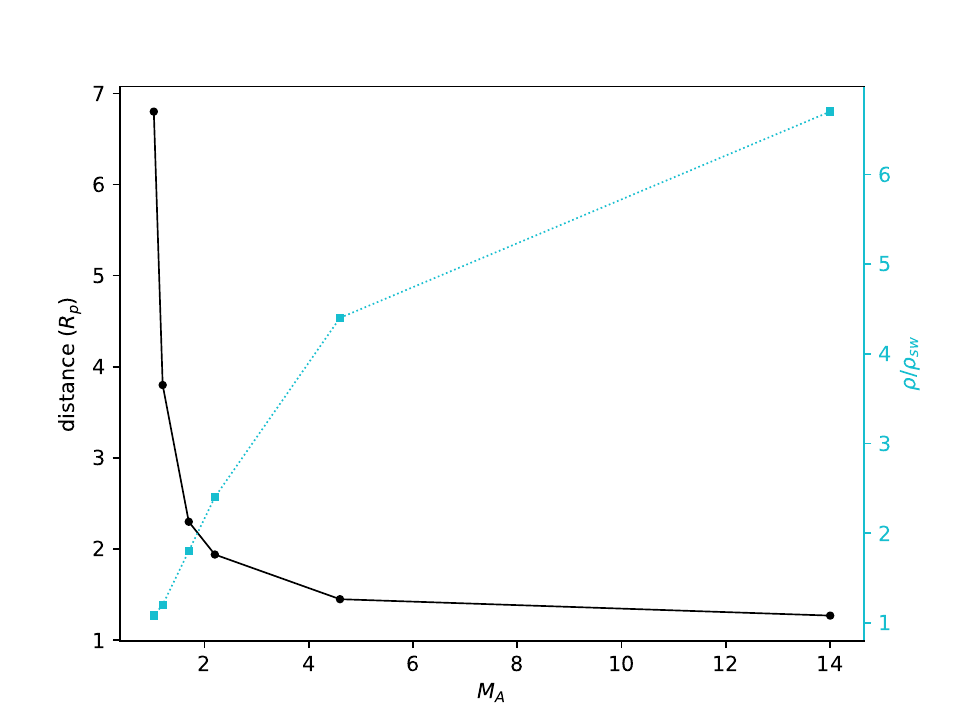}
    \caption{Bow shock location measured from the center of the planet (black solid line) and the density ratio across the the shock at the bow shock location point (dotted cyan line), measured in the line connecting the planetary and stellar, as a function of the $M_A$ number.}
    \label{fig:bs_position}
\end{figure}

\begin{figure}

	\includegraphics[scale=0.58]{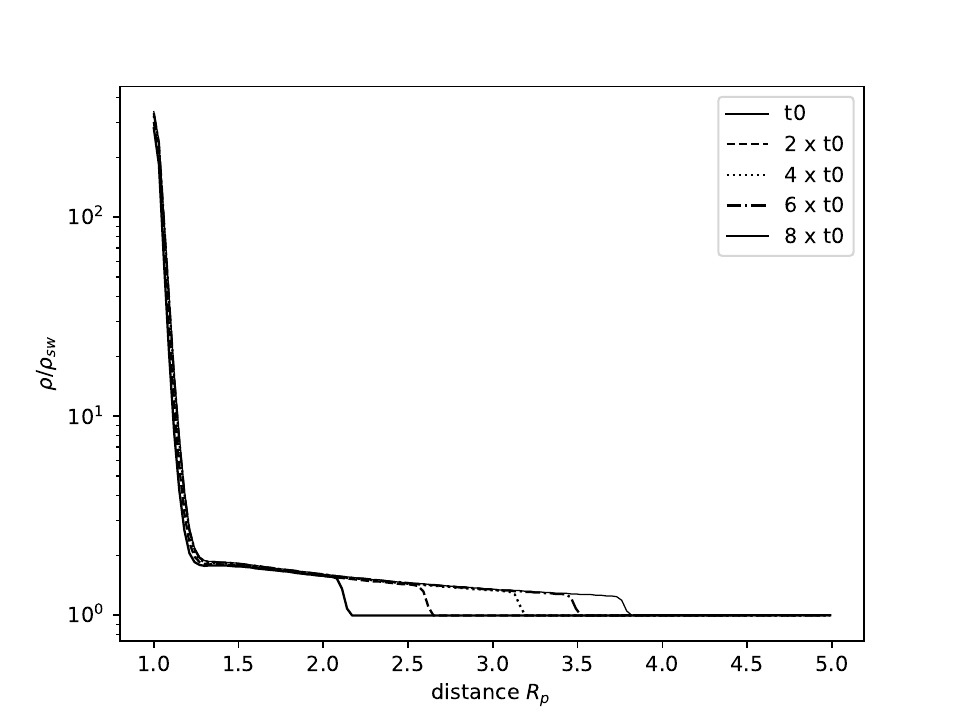}
    \caption{Temporal evolution of the bow shock formed in model M5. The density contrast profile is shown at different time steps of the simulation, from a $t_0$ reference time step.}
    \label{fig:shock_evol}
\end{figure}

\begin{figure*}

	\includegraphics[scale=0.57]{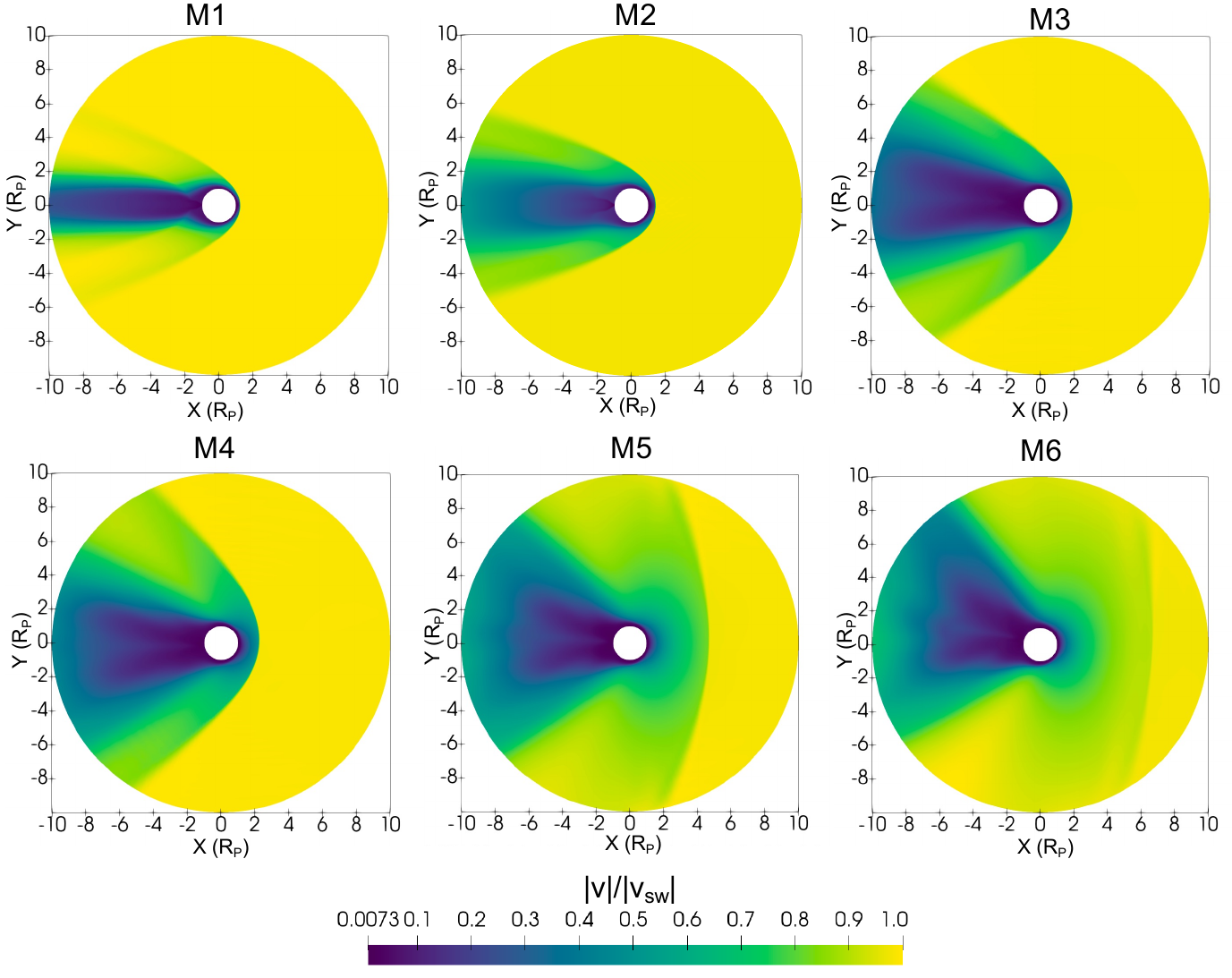}
    \caption{Velocity field magnitude distribution around the planetary obstacle for each simulated model. The velocity magnitude is normalized by the interplanetary velocity value, defined as injected in the outer boundary side of the domain.}
    \label{fig:mosaic_vf}
\end{figure*}

\begin{figure}

	\includegraphics[scale=0.6]{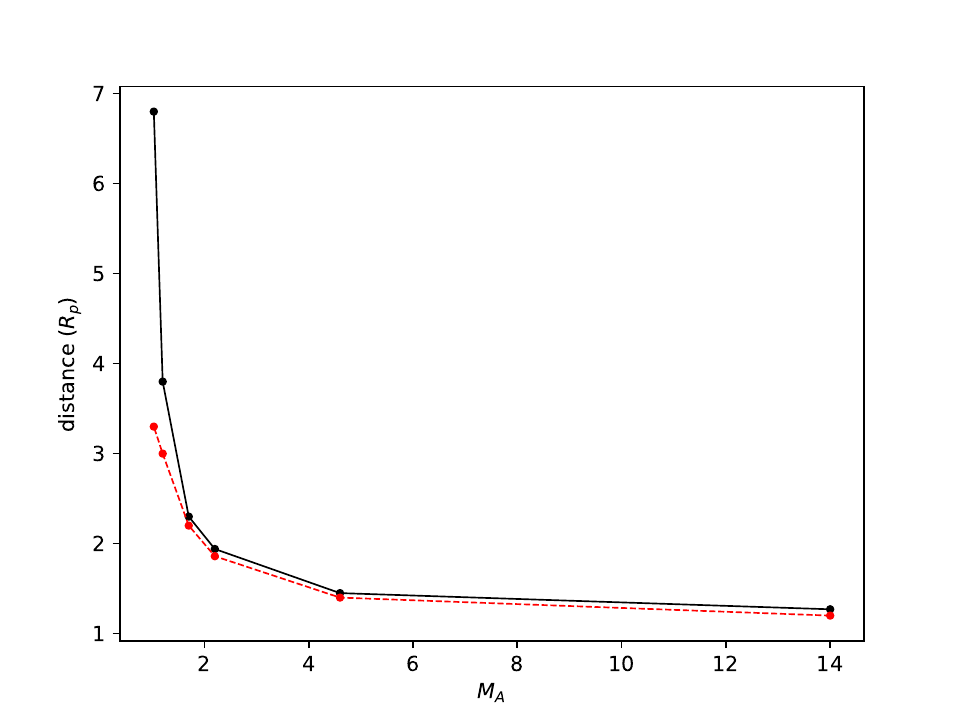}
    \caption{Magnetopause position as function of the alfv\'enic Mach number (red dashed line). The position of the bow shock for each wind model is represented by the black solid line for comparison.}
    \label{fig:mp_position}
\end{figure}

\begin{figure*}
	\includegraphics[scale=0.55]{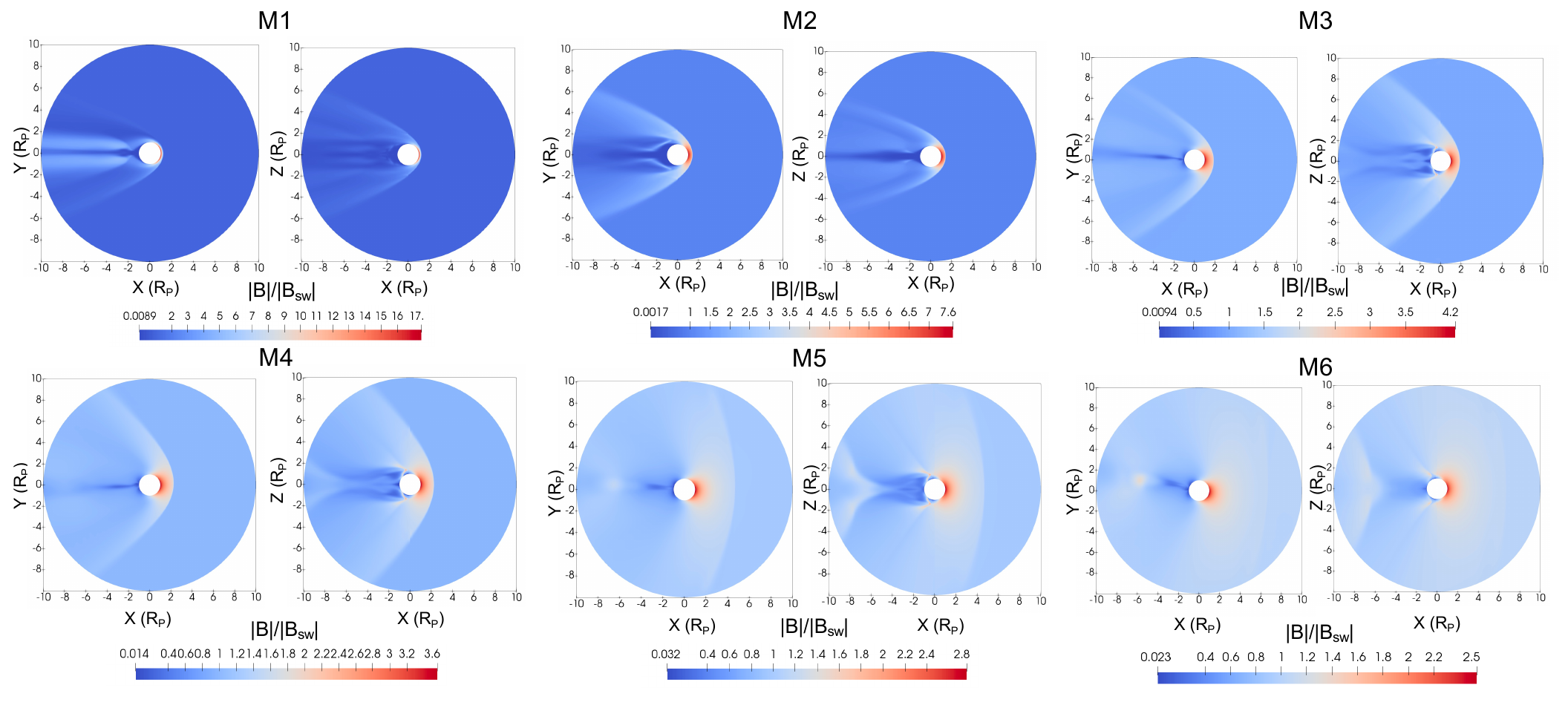}
    \caption{Magnetic filed distribution around the planet in the steady state of the simulation, for each wind considered model. The magnetic field is normalized by the interplanetary magnetic field value of the stellar wind, defined at the right boundary side of the domain. }
    \label{fig:mosaic_mf}
\end{figure*}

\begin{figure*}

	\includegraphics[scale=0.44]{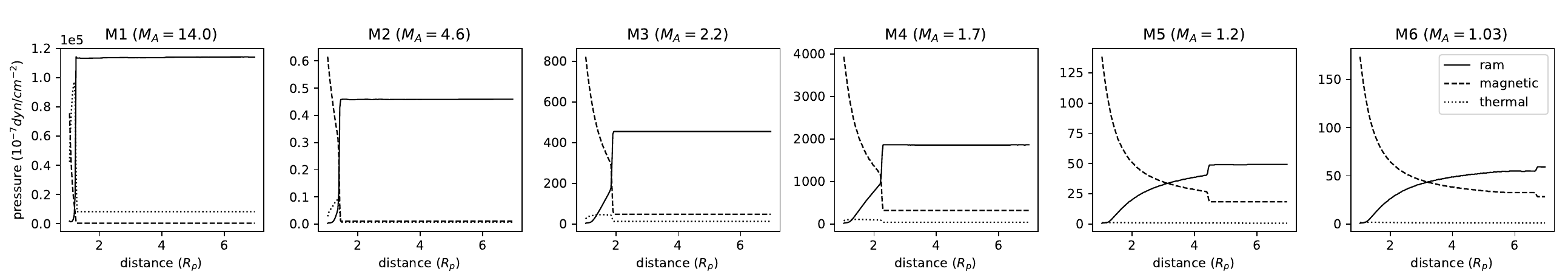}
    \caption{Ram (solid line), magnetic (dashed line) and thermal (dotted line) pressure profiles at Y = 0 along the the star-planet line (X axis). Distances are given in planetary radii, where the planet's surface is located at r= 1.0$R_p$.}
    \label{fig:prs_profiles}
\end{figure*}

We performed a total of six simulation models of the interaction of the superfast-magnetosonic wind of AB Dor with a highly conductive, unmagnetized, tenuous atmosphere planet, for different $M_A$ and $M_{ms}$ parameter winds, as listed in \autoref{tab:table2}.\\ The main feature of the stellar wind-planet interaction in the case of superfast-magnetosonic winds is the formation of a bow shock in front of the planet, characterized by a sudden increase in the local density and plasma temperature. \autoref{fig:mosaic_dens} shows the local density distribution around the planetary obstacle in the XY and XZ planes at the steady state for each model; densities are normalized by the interplanetary density of the wind ($\rho_{sw}$) defined at the outer boundary of the domain. Our simulations show that large bow shocks can be formed around planetary obstacles depending on the coronal properties of the fast-rotating, highly magnetized star. The bow shock position, measured along the sub-stellar line (X axis) from the center of the planet, shows a clearly dependence with the $M_A$ parameter; in the steady state, the steady shock front is found closer to the planetary surface for high-$M_A$ winds (models M1-M4), while it is as far as 6.8 $R_p$ for low-$M_A$ winds (M6). Following the results of our simulations, the bow shock extension is governed by the $M_A$ wind parameter in the case of non-magnetized, highly conductive rocky planets, instead of the $M{ms}$ number, a feature that  is shown comparing models M1 and M2. While model M1 has a $M_A$ number of 14.0 and a $M_{ms}$ number of 3.6, model M2 has a lower $M_A$ number of 4.0 and a higher $M_{ms}$ number of 4.0, finding a slightly more extensive bow shock for model M2. However, the extension of the shock along the Y axis is larger for lower $M_{ms}$ (see XY plane density distributions for models M1 and M2 in \autoref{fig:mosaic_dens}).

As the bow shock extension increases for decreasing $M_A$ winds, it becomes less compressive, decreasing the density ratio between the local stellar wind and the density inside the shock (see \autoref{fig:bs_position}). For model M1, characterized by a high alfv\'enic Mach number $M_A=14.0$, the density measured just behind the shock is 6.7 times the value of the local stellar wind. This density ratio exceeds the value of $\sim$4 expected for purely hydrodynamical winds (Rankine-Hugoniot shock jump conditions). In model M1, the shock is found at a very close distance to the planetary surface, then, the increase of density across the shock is superposed to the local density of the planetary outflow, giving place to this high-density ratio.\\
Planets exposed to low alfv\'enic Mach number wind show extended bow shocks characterized by a low density contrast across the shock, constituting a transition between the super- and sub-alfv\'enic winds, where no bow shock is expected.

All formed shocks follow a characteristic temporal evolution: in the first time steps of the simulation, the bow shock is formed at very close distances from the planetary surface. As the simulation progresses, the shock moves further away towards the star, decelerating until it reaches its final position at the steady state (\autoref{fig:shock_evol}). The density ratio $\rho/\rho_{sw}$ decreases as it moves away from the planetary surface until it reaches its stationary state. As expected, the time to reach this state is longer for more extended the bow shocks.

As a consequence of the deflection of the stellar wind around the obstacle, a wake is formed behind the planet. This wake is characterized in all studied cases by an accumulation of plasma just behind the planet and low flow velocities (\autoref{fig:mosaic_vf}). For each model, the region is shielded from direct interaction with the stellar wind, resulting in density ratios increasing up to three orders of magnitude higher than the local stellar wind density. This is a well-known structure resulting from the wind-planet interaction \citep[e.g. \citetalias{2021MNRAS.502.6170C}, ][]{2015ApJ...806...41C}. In the equatorial (XY) plane, the maximum confinement of density is found at Y=0, just behind the planet. This high-density region is surrounded by low density (0.1$\rho/\rho_{sw}$) areas in all cases. The morphology of the planetary tail-shaped wake is also dependent on the parameters of the stellar wind. This region is more confined to the Y=0 direction in the case of high alfv\'enic Mach number winds, showing more extensive wakes in the vertical direction for low-Mach number winds. Slightly asymmetric tails are found in the XY plane for low-$M_A$ models. This region appears to be more turbulent and unstable in comparison to higher-$M_A$ models, as low-$M_A$ winds can be naturally identified as the transition between super-alfv\'enic and sub-alf\'enic winds, where extensive wakes are found. A similar wake morphology is found for all models in the XZ plane, showing a confinement of density towards the equator. The exception is model M6, where the maximum density is found along the surface of the planet, due to the asymmetry found in the density distribution along the tail.\\

Due to the high conductivity of the planetary surface, the magnetic field strongly piles-up in front of the planet, resulting in a general increase of the magnetic pressure inside the shock. The strong pile-up of the interplanetary magnetic filed in front of the planet leads to the formation of an induced magnetosphere in all studied cases. An induced magnetosphere can be defined as the region inside the shock where the magnetic pressure dominates the rest of pressure contributions \citep{2004AdSpR..33.1905L,2007Natur.450..650B,2008P&SS...56..796K}. This region is limited by two boundaries. The outer boundary, or the so called magnetopause, can be defined as the surface where the magnetic pressure equals the dynamic pressure of the stellar wind. The second boundary, often called the ionopause, is defined by the presence of a conductive medium, in this case the planetary surface. The extent of the magnetopause increases for decreasing Mach numbers (see \autoref{fig:mp_position}), leading to extensive magnetospheres in the case of low-$M_A$ winds, following a similar trend to the observed in the location of the bow shock. In the case of high-Alfvén Mach numbers (models M1 to M4), the magnetopause appears slightly below the bow shock. The relative position between the magnetopause and the bow shock is increased for lower alfvénic numbers, where the magnetopause is found at 0.8 $R_p$ and 3.5 $R_p$ from the shock position for models M5 and M6, respectively. 

Inside the induced magnetosphere, the highest values of the magnetic field are found close to the planetary boundary, decreasing at further distances from the planet, as shown in \autoref{fig:mosaic_mf}. Again, a clear dependence with $M_A$ is found: the magnetic field inside the shock is higher for higher $M_A$ winds, due to the more compressive magnetosonic wave in those cases. In the case of a wind with $M_A$=14.0, the magnetic field is amplified in the induced magnetosphere up to 17 times the local interplanetary magnetic field. This magnetic field enhancement decreases to 2.5 times the local stellar wind magnetic field in the case of an extremely low-$M_A$ wind number of 1.03 (model M6). \\Even though the orientation of the magnetic field is considered in all models, that is, considering both radial and azimuthal components, no magnetic field asymmetries are found in any model, not even in M1 model, where the radial and azimuthal magnetic field components are comparable. The topology of the formed induced magnetospheres is governed in the studied cases by the ram pressure of the wind, as the magnetic field distribution shows a clear symmetry along the X axis, coincident with the averaged velocity field orientation. \\

The dependency of the terminal shock position and the magnetosphere extension with the alfv\'enic Mach number can be translated in a first approximation into the pressure ratio between the dynamic, $p_{ram}$ and magnetic, $p_{mag}$ contributions of the stellar wind, as the alfv\'enic Mach number can be expressed as:
\begin{equation*}
    M_A = \frac{|\textbf{v}|}{|\textbf{A}|} = \sqrt{0.5\frac{p_{ram}}{p_{mag}}}.
\end{equation*}
Outside the shock, the ram pressure of the wind dominates over the rest of pressure contributions (see \autoref{fig:prs_profiles}), as $M_A$>1 for all simulated cases. The ram pressure drops across the shock, while the magnetic pressure increases due to the formation of the induced magnetosphere. In the case of high-$M_A$ winds, the ram pressure can be as three orders of magnitude higher than the magnetic pressure, pushing the formed shock towards the planet. Even though the shock is more compressing on those cases, heating the plasma in the vicinity of the planet and increasing the local density and magnetic field behind the shock, the dynamics of the shock is governed by the ram pressure of the stellar wind, leading to closer shocks in front of the planet. In the case of low-$M_A$ winds, ram and magnetic pressures become comparable, following the last equation. As the magnetic field increases through the shock, the magnetic pressure exerted by the induced magnetosphere pushes the shock outwards, leading to more extensive bow shocks.

\section{Summary and Conclusions}\label{sec:conclusions}

In the present study, we performed three-dimensional MHD numerical simulations to predict and characterize the structures formed in the local environment of poor-atmosphere, Earth-like planets under the influence of superfast-magnetosonic stellar winds of fast-rotating, low-mass stars, focusing in the case of AB Dor, a prototype of magnetically active, fast-rotating star with a rotation period of 0.51 days. For the first time, the properties of the formed planetary bow shocks and induced magnetospheres have been evaluated for a wide range of stellar wind alfvénic Mach numbers, in the case of planets with tenuous atmospheres, as a probable scenario for highly irradiated Earth-like planets orbiting active stars. \\
Stellar winds have been assumed to be mainly accelerated due to the action of magneto-centrifugal forces. In this work, the velocity, magnetic field and stellar wind density at the planetary orbit have been derived using the analytical solution of the 1D Weber \& Davis model. Under this approximation, extremely accelerated winds are expected in the case of AB Dor, also predicted in the simulations performed in \citet{2010ApJ...721...80C} of the coronal structure of AB Dor. Winds in the superfast-magnetosonic regime seem to be a common feature of low-magnetized (2.0-10.0 G) winds. In the case of extremely magnetized stars, winds become fast-magnetosonic at larger distances from the stellar center, around 15 au. The predicted critical points of the Weber \& Davis model should be however regarded carefully, as the complex 3D morphology of the coronal magnetic field and other cited acceleration mechanisms, as the Alfvén wave heating, have not been considered here, and should be included in future work.\\

As an immediate consequence of the interaction of superfast-magnetosonic wind with a steady obstacle, a bow shock is formed in front of the planet. The extension of the bow shock shows a clear dependence with the alfv\'enic Mach number, as expected, following the results obtained in the case of Solar-System planets. Extended bow shocks are found for low-$M_A$ winds, reaching distances as far as $\sim$7.0 $R_p$ for $M_A$ = 1.03 winds. However, due to the low compression of the shock in those cases, the bow shock is characterized by a low density contrast with the ambient stellar wind ($\rho/\rho{sw}$=1.1 for $M_A$ = 1.03). Bow shocks formed around planets orbiting active solar-like stars can be as $\sim$5 times larger than those found around the non-magnetized planets of the Solar System (as a reference, the Venus' bow shock is found at 1.4 Venusian radii for solar maximum conditions). The shock extension is clearly dependent on the formation of an induced magnetosphere, exerting an additional pressure to the shock. An induced magnetosphere is formed in all configurations, due to the strong pile-up of the interplanetary magnetic field above the highly conductive obstacle. The extension of the induced magnetosphere, limited by the magnetopause boundary, is larger for low-$M_A$ winds. Again, as low-$M_A$ winds are less compressive, the magnetic field magnitude inside the shock is lower in comparison to higher-$M_A$ winds.\\

Other stellar and planetary parameters are known to modify the topology and properties of the formed plasma structures in the stellar wind. The presence of a planetary extended atmosphere means an extra obstacle to the wind, producing the deceleration of the wind plasma through collisions, leading to a pile-up of plasma upstream the planet. In \citetalias{2021MNRAS.502.6170C}, we studied the interaction of Earth-like, fully ionized H-rich exosphere of a non-magnetized planet and the superfast-magnetosonic stellar wind of young solar-like stars. Extended bow shocks appear under the influence of lower $M_A$ winds, corresponding in that study to more evolved stars, increasing the shock extension for denser exospheres under the same stellar wind conditions. The presence of an extended atmosphere then expands the bow shock at further distance from the planetary surface. For the youngest star on that work (0.1 Gyr, $M_A$=22), the bow shock is found at very close distances to the planet, even for the densest considered exospheres due to the high $M_A$ number considered on that cases.

As no chemical reactions are considered in this work, mass loading processes due to charge exchange and photoionization are neglected. Mass loading may contribute to increase the ion population in the vicinity of the planet, producing an strong deceleration of the stellar wind plasma that could enhance the magnetic field in front of the planet, producing more extensive induced magnetospheres. In \citet{2013JGRA..118..321M}, they compared their results in the case with and without mass loading processes, for the case of Venus. They found bow shock formation slightly closer to planetary surface in comparison to the mass load case. Then, initially considering a neutral component in the planetary thin atmosphere will contribute to the mass loading, increasing the extension of the shock. 

Other parameters that could enhance the extension of the bow shock and the induced magnetosphere, such as the influence of EUV radiation, are not addressed in this study. EUV radiation pressure is known also to modify the topology and location of the shock, as demonstrated for Mars and Venus. The influence of the stellar radiation is crucial in the study of planetary structures, and we will address the contribution of radiation in future work.

Simulated planets in this work are supposed to remain in the same (super-alfvénic) regime along its orbit, not considering temporal evolution of the stellar wind conditions. Moreover, due to the non uniform distribution of Alfvén critical surfaces around the star, as predicted in more complex 3D stellar wind simulations, the planet may experience wind conditions changing from sub- to super-alfvénic regime \citep{2015ApJ...806...41C}. In addition, transient events such as coronal mass ejections or an enhanced stellar wind activity change the wind magnetic conditions, modifying the characteristic Mach numbers of the wind, and leading the changes in the morphology of the shock and the induced magnetospheres. However, our parametric study, covering a wide range of alfvénic Mach numbers for the wind, is able to collect different conditions of solar wind activity, changing the alfvénic Mach number from higher values (minimum activity) to lower values (maximum activity), following the solar wind observations.

According to the results of the present work, extended MHD structures are found even in the case of unmagnetized, poor-atmosphere planets. However, the prospects of detection are scarce: the density ratio of these structures in relation to the ambient wind is extremely low, especially in the case of the most extended structures, only showing a considerable increase of density flux in the wake of the planet. This leads to low possibilities of structure detection in the case of these bare planets orbiting active solar-like stars. Other evolutionary stages, where an atmosphere is still present, should be analysed in order to obtain detectable MHD structures around these Earth-like planets, shedding light on the early evolution of potential life-hosting planets.

\section*{Acknowledgments}
We thank the referee's insightful comments which have helped to improve significantly the manuscript. We are also very grateful for the invaluable help of Andrea Mignone and Jacobo Varela in the numerous aspects of the simulations carried out and presented in this paper. This work has been partially financed by the Ministry of Science and Innovation through grants: ESP2017-87813-R and PID2020-116726RB-I00. 


\section*{Data Availability}
The data underlying this article are available in the article and in its online supplementary material.



\bibliographystyle{mnras}
\bibliography{my_bibliography} 


\bsp	
\label{lastpage}
\end{document}